\documentclass[aps,prl,twocolumn,noshowpacs]{revtex4-1}
\usepackage{amsmath}
\usepackage{color}
\usepackage{graphicx}
\usepackage{epstopdf}

\begin{document}

\title{Nucleation at the phase transition near 40~$^{\circ }$C in MnAs nanodisks}
\author{B.~Jenichen}
\email{jen@pdi-berlin.de}
\author{Y.~Takagaki}
\author{K. H.~Ploog}
\affiliation{Paul-Drude-Institut f\"{u}r Festk\"{o}rperelektronik,
Hausvogteiplatz 5--7, D-10117 Berlin, Germany}
\author{N.~Darowski}
\author{R.~Feyerherm}
\author{I.~Zizak}
\affiliation{Hahn-Meitner-Institut Berlin GmbH, Glienicker Strasse
100, D-14109 Berlin, Germany}
\date{\today}

\begin{abstract}
The phase transition near 40~$^{\circ }$C of both as-grown thin
epitaxial MnAs films prepared by molecular beam epitaxy on
GaAs(001) and nanometer-scale disks fabricated  from the same
films is studied. The disks are found to exhibit a pronounced
hysteresis in the temperature curve of the phase composition. In
contrast, supercooling and overheating take place far less in the
samples of continuous layers. These phenomena are explained in
terms of the necessary formation of nuclei of the other phase in
each of the disks independent from each other. The influence of
the elastic strains in the disks is reduced considerably.
\end{abstract}


\maketitle

%%,07.85.Jy diffractometer,81.15.Hi MBE, 61.10.Nz X-ray diffraction
%%61.10.Eq X-ray scattering,68.35.Bs surface structure and topography
%% 68.55.Nq Thin film structure and morphology, composition and phase identification
%% 75.50 Cc Other ferromagnetic materials and alloys

%%\section{Introduction}
Manganese arsenide on GaAs is a promising materials combination
for spintronic applications based on spin injection \cite{ram02}.
MnAs is ferromagnetic at room temperature and has a large carrier
spin polarization. It can serve as a source of spin polarized
electrons. Furthermore MnAs may be applied for sensors and
actuators thanks to its magnetoelastic response
\cite{tschernenko99}. The room temperature ferromagnetic
$\alpha$-phase is metallic and crystallizes in the hexagonal
NiAs($B8_1$) structure. Near 40~$^{\circ }$C MnAs transforms into
the orthorhombic $\beta$-phase exhibiting the MnP(B31) structure.
The temperature dependence of the magnetization near the
transition was investigated in \cite{tschernenko99,govor86}, and a
20~K wide hysteresis loop  was observed. At this first order
structural phase transition a significant change in the lattice
parameter $a$ is found, which amounts to $\approx $1.2\%
\cite{willi54,wilson64}.

\begin{figure}[tbp]
\includegraphics[width=8.5cm]{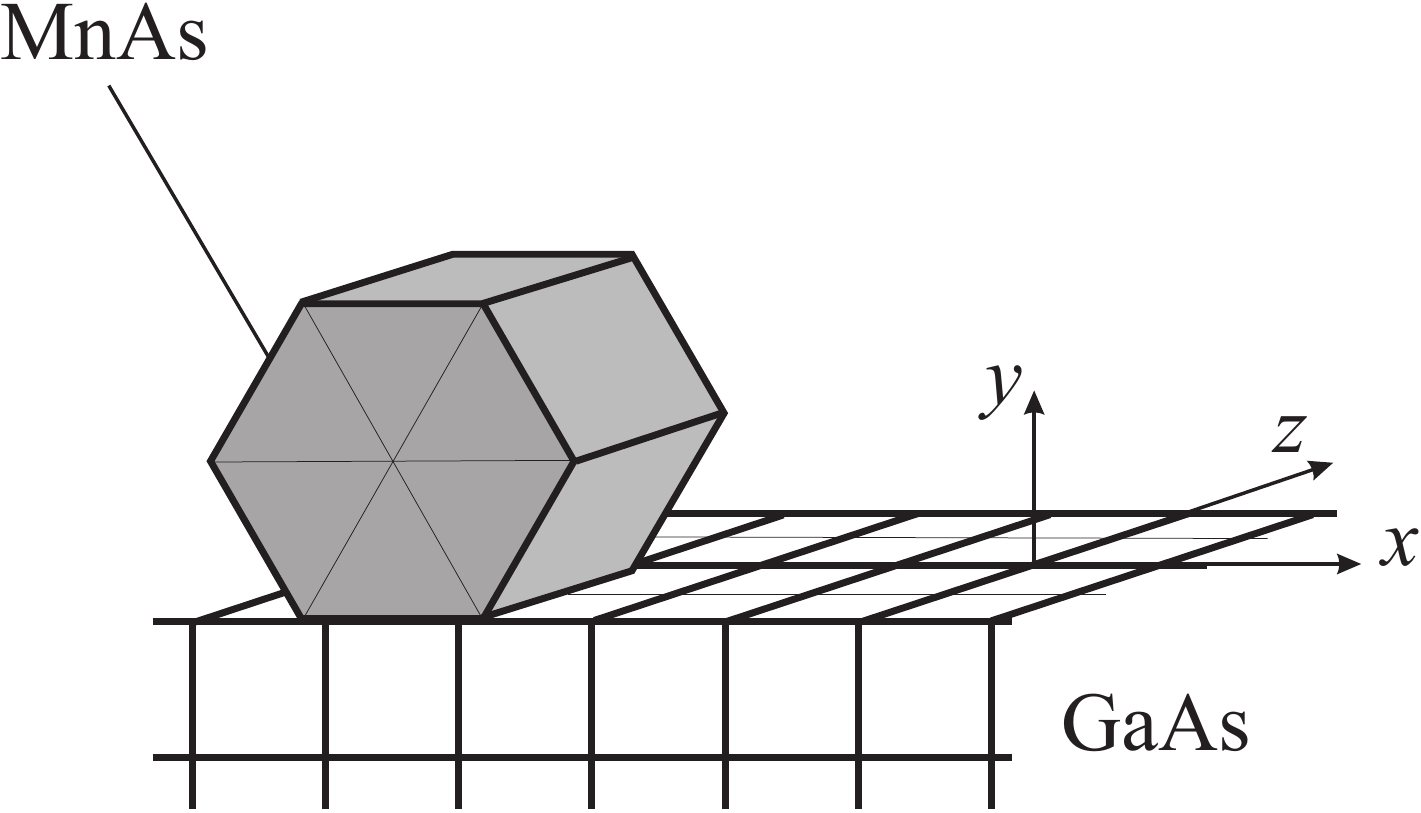}
\caption{Schematic view of the epitaxial relationship of MnAs on GaAs(001).\vspace{5cm}}
\label{fig:epitaxyMnAs}
\end{figure}

During epitaxy \cite{tanaka94,schippan99} the MnAs(1\={1}00) film
on GaAs(001) is attached by the side facet of the hexagonal unit
cell (Fig.\ \ref{fig:epitaxyMnAs}) , so that the $c$ lattice
direction MnAs[0001] is parallel to GaAs[1\={1}0]. The
deformations in the layer lead to the phenomenon of phase
coexistence, i.e. the phase content $\xi$ does not change abruptly
between zero and unity at a certain temperature as expected from
the Gibbs phase rule. Two coexisting phases are found, on the
contrary, in a wide temperature range \cite{kag00}. Elastic
domains of both phases form a periodic stripe pattern \cite{kag02}
in a self organized way. The domain period amounts to
approximately five times the film thickness \cite{plake02,jeni04}.
Application of hydrostatic pressure \cite{bean62,menyuk69} or
biaxial stress \cite{iikawa05} has a considerable influence on the
transition temperatures.

The phase transition can be affected further by imposing
artificial constraints on the stripe pattern. Significant effects
are expected from a lateral confinement when a film is patterned
to small disks. Such disks with smaller sizes than the widths of
the elastic domains enable elastic relaxation of the laterally
periodic stresses accumulated inside the epitaxial layer. The
tight restriction of the MnAs lattice along the interface is then
released. As a consequence the formation of elastic domains in
MnAs nanodisks seems to be no longer energetically favourable. The
distribution of magnetic domains in such MnAs disks was
investigated in \cite{taka05}. The aim of the present work is to
study in more detail the influence of such a lateral structuring
on the phase coexistence of $\alpha$ and $\beta$MnAs. We
investigate the temperature dependence of the phase composition in
epitaxial MnAs films prior to and following the artificial
modification using microfabrication technologies.

\begin{figure}[tbp]
\includegraphics[width=8.5cm]{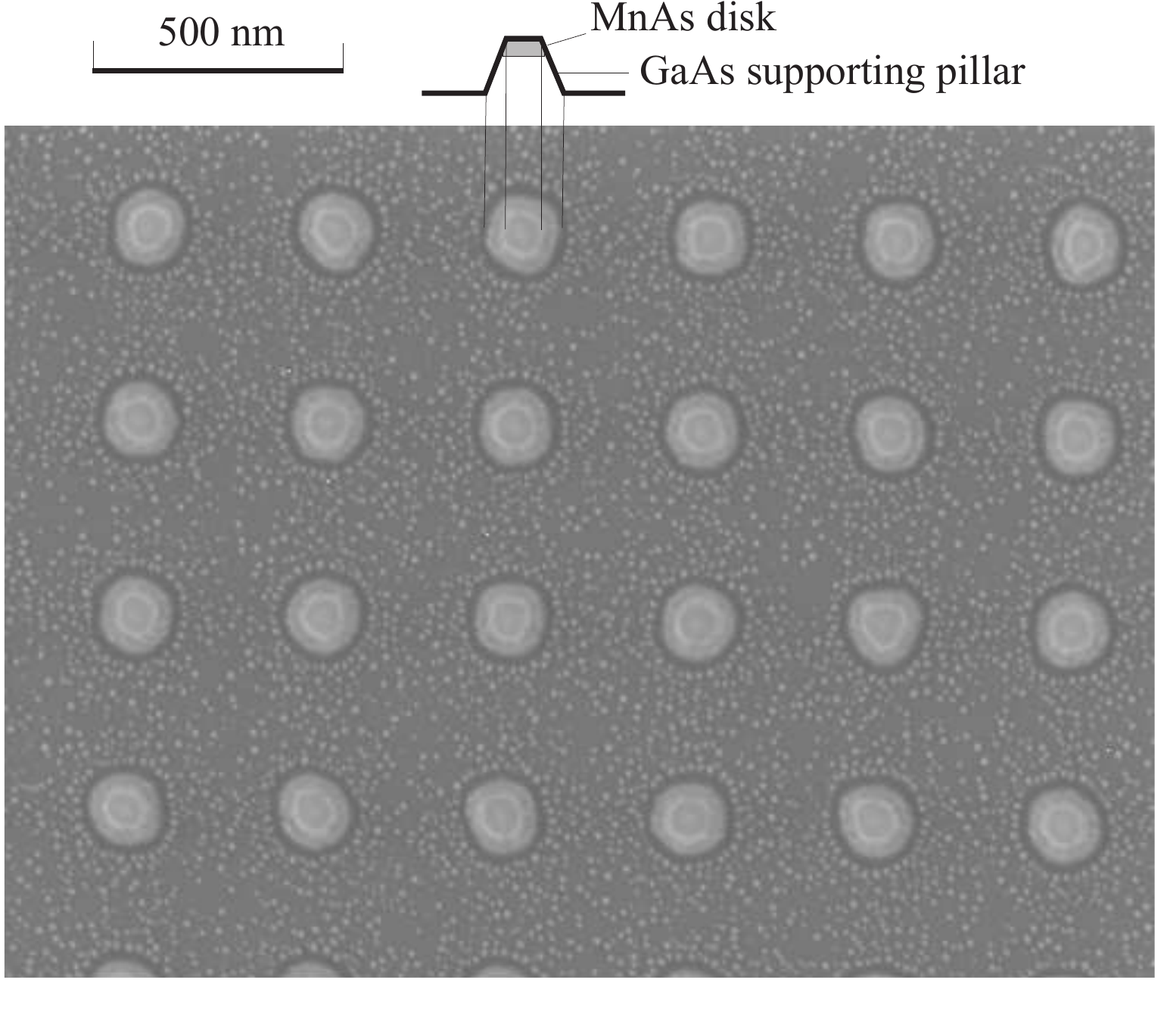}
\caption{Scanning electron micrograph of MnAs disks on GaAs supporting
pillars. The diameter of the disks is below 100~nm, i.e. below the
size of the elastic domains in the original MnAs epitaxial layer.\vspace{5cm}} \label{fig:Dsem}
\end{figure}

%%\section{Experimental}
The MnAs layers were grown by solid source molecular beam epitaxy
(MBE) as described elsewhere \cite{schippan99,kaest00,daw04}.
%%For a layer thickness of 50~nm we can expect a periodicity
%%$\Lambda_{d}$ of the lateral domains of about 250~nm \cite{kag02}.
%%In the epitaxial layer even the $\alpha$ phase is slightly
%%orthorhombic and in principle the hexagonal symmetry is lost
%%\cite{iikawa05}. These deformations are small and we keep the
%%hexagonal indexing for the $\alpha$ phase as usual in epitaxial
%%systems.
The nanostructuring was carried out using electron beam
lithography and Ar ion milling. The resulting disks were assembled
in the form of a square array as shown in the scanning electron
micrograph (SEM) in Fig.~\ref{fig:Dsem}.  In this sample, the
diameter of the disks is smaller than 100~nm, i.e. well below the
equilibrium size of the elastic domains in the original continuous
MnAs layer \footnote[1]{We take into account that the disks have a
conical shape and take as the disk diameter the size of the
smaller ring shaped contrast. The base of every disk is etched
deeply into the GaAs material and has a larger diameter.}.
Temperature dependent synchrotron x-ray diffraction experiments
were performed at the MAGS beamline at the BESSY storage ring
using a Si($111$) double crystal monochromator and 8 keV
radiation. A (3+3) circle diffractometer equipped with a special
cryostat was employed for the measurements. Preliminary
experiments were performed at a similar diffractometer of the
KMC~2 beamline at BESSY. In addition we performed laboratory
experiments using a Panalytical X'Pert System with Ge ($220$)
hybrid monochromator and Ge ($220$) analyzer crystal.

\begin{figure}[tbp]
\includegraphics[width=8.5cm]{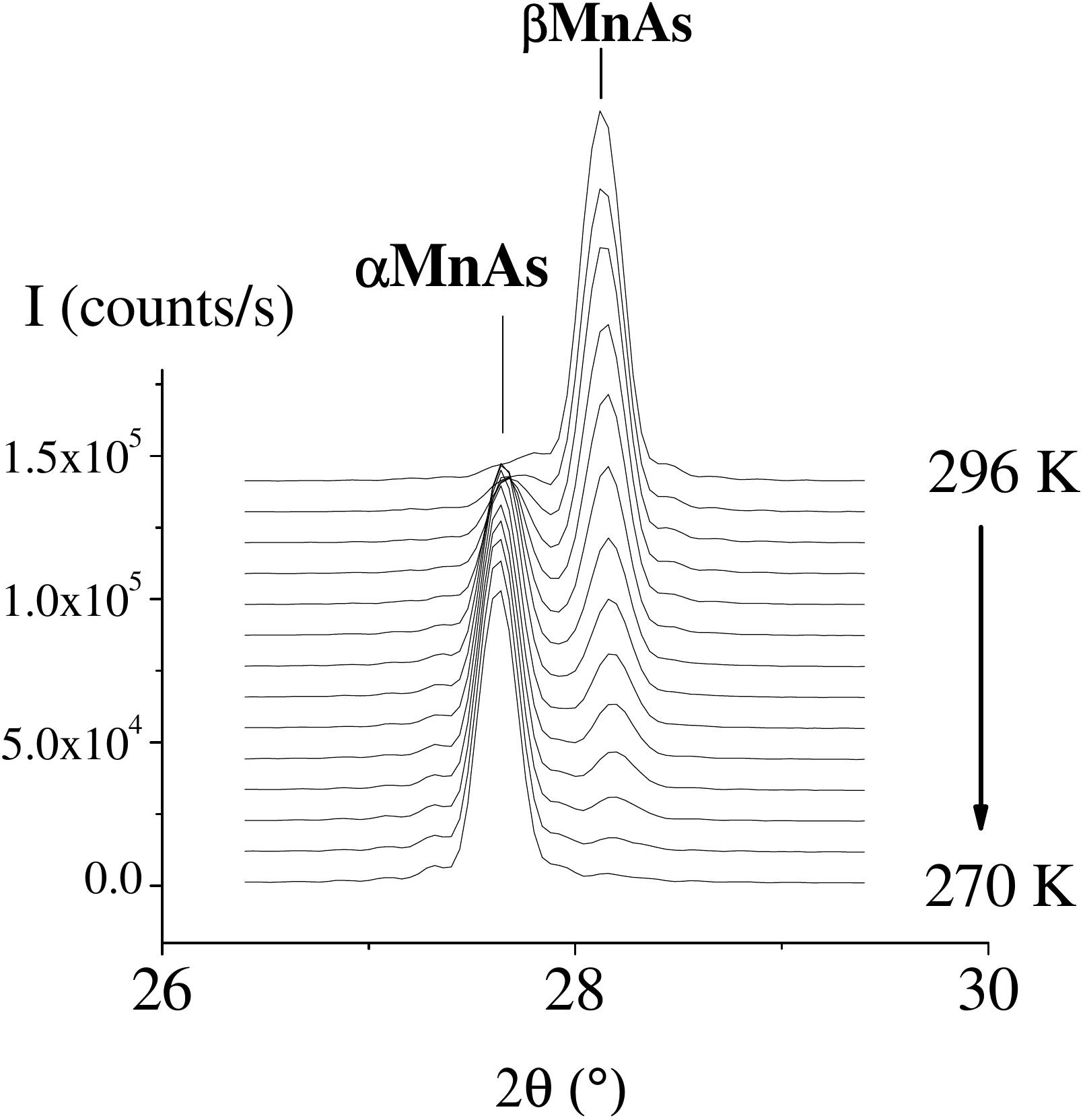}
\caption{X-ray reflections associated with $\alpha $MnAs($\bar{1}100$) and
$\beta $MnAs(020) of the MnAs disks obtained upon cooling the
sample from 296 K to 270 K in steps of 2 K. The curves are shifted
vertically for clarity. \vspace{5cm}  } \label{fig:transiton}
\end{figure}

\begin{figure}[tbp]
\includegraphics[width=8.5cm]{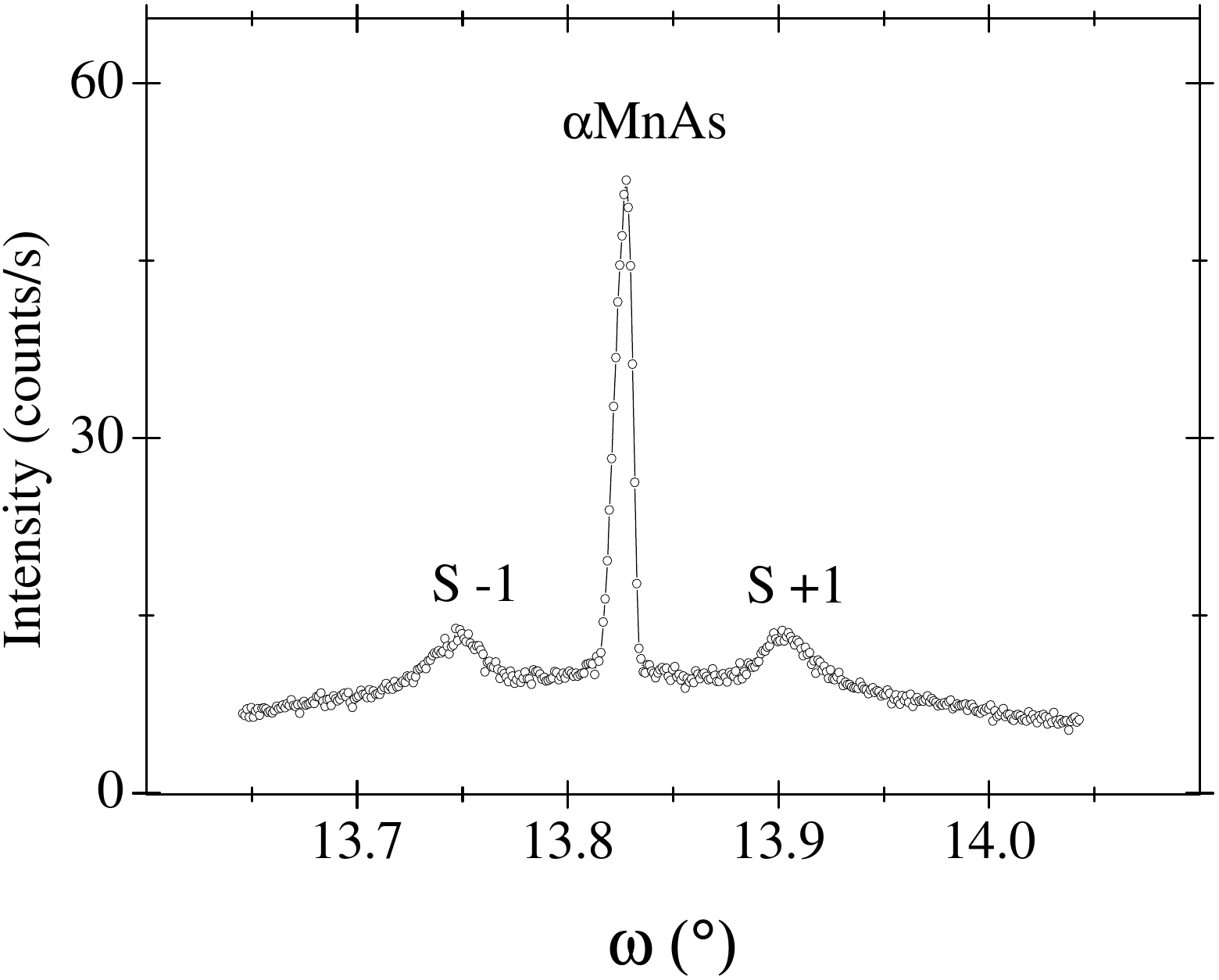}
\caption{X-ray triple crystal $\omega $-scan of the continuous layer
measured at room temperature 27~$^{\circ }$C near the reflection
 $\alpha $MnAs($\bar{1}100$) revealing additional satellite reflections
 (marked by S -1 and S+1) due to a lateral periodicity of the elastic
 domains of 247~nm.\vspace{5cm} } \label{fig:Lsat}
\end{figure}

%%\section{Results}
The phase contents of the MnAs samples were obtained from the
ratio of the integrated intensities of the corresponding
$\alpha$MnAs and $\beta $MnAs reflections measured in symmetrical
$\omega$/2$\theta$-scans (Fig.\ \ref{fig:transiton}). The
($\bar{3}300$) and (060) or the ($\bar{1}100$) and (020) were
analyzed \cite {jeni02,jeni04}, and the layer reflections were
fitted by Gaussian curves. The intensity ratio changes with
temperature in the phase coexistence range \cite{kag00}. The
samples reached their equilibrium composition almost immediately
after a certain temperature had been set, i.e. the relaxation
times are significantly small. Samples consisting of MnAs disks
having various diameters on the GaAs substrate were compared to
their parent unstructured samples. Here we demonstrate the results
from the smallest disks shown in Fig.\ \ref{fig:Dsem}. The lateral
period of the domain structure of the original MnAs epitaxial
layer can be obtained from the distance between satellite maxima
$\Delta\omega_S$ in the x-ray triple crystal $\omega$-scan (Fig.\
\ref{fig:Lsat}). The period $\Lambda_{d}$ is calculated from the
formula $\Lambda_{d}$=2$\pi/(\Delta
Q_x)$=$\lambda$/(2$\Delta\omega_S \sin\Theta_B$), where $\Delta
Q_x$ is the distance of the satellite maxima in reciprocal space,
$\lambda$ is the x-ray wavelength, and $\Theta_B$ is the Bragg
angle \cite{jeni93}. The $x$-axis is defined to be perpendicular
to the $c$-direction of MnAs and parallel to the interface (see
Fig.\ \ref{fig:epitaxyMnAs}). The angular distance of the
satellite maxima in Fig.\ \ref{fig:Lsat} measured at room
temperature yields an average lateral period of the domain
structure of 247~nm. The thickness of the original MnAs film was
determined to be 38~nm using x-ray reflectivity measurements
\cite{jeni04}. The equilibrium domain period is thus estimated to
be 190~nm \cite{kag02}. As the diameters of the smallest disks are
sufficiently small ($\approx$80~nm) only one elastic phase domain
exists in an individual disk, which was confirmed at room
temperature using magnetic force microscopy \cite{taka05}.

\begin{figure}[b]
\includegraphics[width=8.5cm]{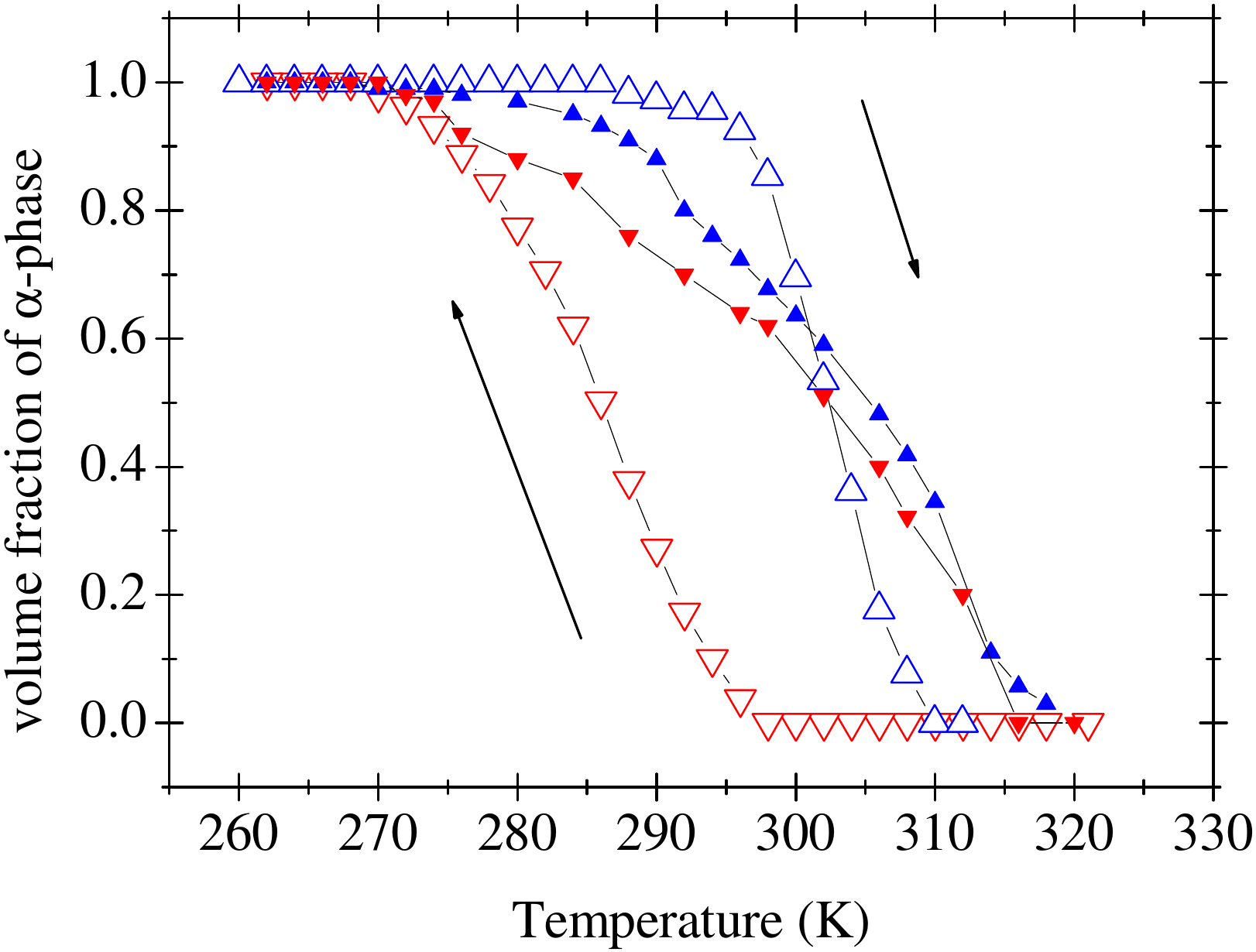}
\caption{(Color online) Temperature dependence of the volume fraction $\xi$
of the MnAs $\alpha$-phase illustrating the coexistence of the two
phases in the MnAs layer (full symbols) and the MnAs disk (hollow
symbols) system. Upwards (downwards) directed triangles correspond
to the heating (cooling) curve. The changes of the composition
with temperature are more steep in the disk system than in the
continuous layer system.  The range of phase coexistence in the
layer system is as large as 45 degrees and only a very small
hysteresis is observed.The disk system changes from one phase to
the other within 15 to 20 degrees and a strong hysteresis (width
$20\pm5~K$) of the temperature curve is found.
\vspace{5cm} } \label{fig:Dcomposition}
\end{figure}

The full triangles in Fig. \ref {fig:Dcomposition} show the
temperature dependence of the phase content $\xi$ of $\alpha$MnAs
in the unpatterned continuous epitaxial layer. As reported in
\cite{kag00} the heating and  cooling curves roughly coincide and
hence the temperature hysteresis in the range of phase coexistence
is negligible. In the present sample this range extends quite
broad between 270 and 315~K. The overall phase coexistence range
amounts to 45~K. In the vicinity of the transition temperature of
315~K the $\alpha$MnAs content rises from zero almost linearly.
When lowering the temperature further the rise of the phase
content weakens, and the content gradually reaches  the saturation
level at unity. The temperature dependence of the phase content
$\xi$ in the small MnAs disks is also shown in Fig.
\ref{fig:Dcomposition} (hollow symbols). When cooling down the
MnAs disks, $\alpha$MnAs first emerges in the disks only at a
temperature as low as 298~K. We observe a significant supercooling
of the disks, i.e. all of them remain to be in the $\beta$-phase.
Subsequently, the $\alpha$MnAs content rises with further cooling
 until all the disks are transformed to the
$\alpha$-phase at 270~K. Once the $\alpha$-phase had been realized
entirely in all the disks, the sample was heated. Similar to the
cooling case the temperature was as high as 285~K when the disks
began transforming into the $\beta$-phase. Therefore, a
significant extent of overheating takes place in the disks in
contrast to the continuous MnAs layer.
%%\section{Discussion}
The behavior of the MnAs nanodisks at the first order phase
transition is similar to that of bulk MnAs
\cite{tschernenko99,govor86}. The same widths of the hysteresis in
the temperature dependencies of the magnetization in bulk MnAs
\cite{tschernenko99} and of the phase content  $\xi(T)$ in the
MnAs disk ensemble are found.  Moreover no phase coexistence takes
place in the individual disks \cite{taka05}, indicating that the
contribution of the elastic deformations during the phase
transition is reduced considerably in the disk system.
Nevertheless, the phase transition in the disk ensemble does not
occur abruptly at a certain temperature. The slope of the
temperature curve $d \xi/ d T$ has increased only by a factor of
2-3 compared to that of the layer curve. The fact that the
experimental disks are not perfectly identical due to small
fluctuations in their sizes and shapes and the random presence of
defects may be responsible for the finite temperature window at
the phase transition. The strong temperature hysteresis observed
in the experiment (Fig. \ref{fig:Dcomposition}) manifests the
supersaturation in individual disks. The development of the other
phase is retarded by a barrier, the energy $\Delta f^*$ of
formation of critical nuclei of the other phase \footnote[2]{In
the isotropic approximation we find from the width of the
hysteresis loop in the temperature curve of the MnAs disks $\Delta
f^* = (4/3)\pi r_c^2 \sigma_b$, where $\sigma_b\approx 8~m J/m^2$
is the interface tension of the phase boundary and
$r_c\approx9~nm$ is the critical radius of the nuclei of the other
phase \cite{landau75, bohm88}.}.  This energy barrier seems to be
connected mainly to the energy of the created phase boundary as in
the case of bulk MnAs. The influence of the elastic energy, which
was most important in the case of MnAs epitaxial films, is
reduced.

%%\section{Summary}

In conclusion, we compared the first-order phase transition in
MBE-grown MnAs films on GaAs and in nanodisks prepared from the
same MnAs films.
%%The films exhibit a large range of phase
%%coexistence ($\sim45$ degrees).
%% but negligible hysteresis of the
%%temperature dependence of the phase content.
The disks show supercooling (overheating) effects and as a
consequence a pronounced hysteresis in the temperature dependence
like in bulk MnAs. A stable nucleus of the other phase is required
in each of the disks, since the individual disks are independent
from each other. On the contrary, fewer nuclei are needed in the
continuous layer as they can grow larger to fill the whole layer
without restriction.
%%The influence of the elastic deformations in the
%%disks is
%%reduced compared to the sample with continuous film.
\section{Acknowledgement}

The authors thank E.~Dudzik, E.~Wiebecke, C.~Herrmann,
V.~M.~Kaganer, L.~D\"aweritz,  and A.~Erko for their support and
for helpful discussions.

%\begin{table*}[tbp]
%\caption{Structural parameters and the low temperature
%resistivities of the samples under investigation}
%\begin{ruledtabular}
%\begin{tabular}{|c|c|c|c|c|c|c|c|c|}
%       & Si cell     & Si content  & layer     &        & \multicolumn{2}{c|}{order parameters} & relaxation & resistivity           \\
%sample  & temperature & in Fe$_3$Si & thickness & misfit & $\alpha$       & $\beta$             &  $z$       &        \\
%        &($^{\circ}$C)& (\%)        & (nm)      & (\%)   & to Fe(B)       & to Fe(A,C)          & ({\AA})    & ($\mu\Omega$~cm) \\ \hline
%1       & 1395        & 25.5        & 32.6      & -0.06  & 0              & 0.3                  & 0.14       & 23               \\
%2       & 1390        & 22.5        & 28.4      & 0.15   & 0.25           & 0.28                 & 0.06       & 50               \\
%3       & 1375        & 16.5        & 28.2      & 0.62   & 0.25           & 0.5                 & 0.08       & 103              \\
%\end{tabular}
%\end{ruledtabular}
%\label{tab:results}
%\end{table*}

\newpage
%\section{References}
\bibliographystyle{apsrev}

%%\bibliography{Zitate}

\begin{thebibliography}{21}
\expandafter\ifx\csname
natexlab\endcsname\relax\def\natexlab#1{#1}\fi
\expandafter\ifx\csname bibnamefont\endcsname\relax
  \def\bibnamefont#1{#1}\fi
\expandafter\ifx\csname bibfnamefont\endcsname\relax
  \def\bibfnamefont#1{#1}\fi
\expandafter\ifx\csname citenamefont\endcsname\relax
  \def\citenamefont#1{#1}\fi
\expandafter\ifx\csname url\endcsname\relax
  \def\url#1{\texttt{#1}}\fi
\expandafter\ifx\csname
urlprefix\endcsname\relax\def\urlprefix{URL }\fi
\providecommand{\bibinfo}[2]{#2}
\providecommand{\eprint}[2][]{\url{#2}}

\bibitem[{\citenamefont{Ramsteiner et~al.}(2002)\citenamefont{Ramsteiner, Hao,
  Kawahrazuka, Zhu, K\"astner, Hey, D\"aweritz, Grahn, and Ploog}}]{ram02}
\bibinfo{author}{\bibfnamefont{M.}~\bibnamefont{Ramsteiner}},
  \bibinfo{author}{\bibfnamefont{H.~J.} \bibnamefont{Hao}},
  \bibinfo{author}{\bibfnamefont{A.}~\bibnamefont{Kawahrazuka}},
  \bibinfo{author}{\bibfnamefont{H.~J.} \bibnamefont{Zhu}},
  \bibinfo{author}{\bibfnamefont{M.}~\bibnamefont{K\"astner}},
  \bibinfo{author}{\bibfnamefont{R.}~\bibnamefont{Hey}},
  \bibinfo{author}{\bibfnamefont{L.}~\bibnamefont{D\"aweritz}},
  \bibinfo{author}{\bibfnamefont{H.~T.} \bibnamefont{Grahn}}, \bibnamefont{and}
  \bibinfo{author}{\bibfnamefont{K.~H.} \bibnamefont{Ploog}},
  \bibinfo{journal}{Phys. Rev. B} \textbf{\bibinfo{volume}{66}},
  \bibinfo{pages}{081304} (\bibinfo{year}{2002}).

\bibitem[{\citenamefont{Chernenko et~al.}(1999)\citenamefont{Chernenko, Wee,
  McCormick, and Street}}]{tschernenko99}
\bibinfo{author}{\bibfnamefont{V.~A.} \bibnamefont{Chernenko}},
  \bibinfo{author}{\bibfnamefont{L.}~\bibnamefont{Wee}},
  \bibinfo{author}{\bibfnamefont{P.~G.} \bibnamefont{McCormick}},
  \bibnamefont{and} \bibinfo{author}{\bibfnamefont{R.}~\bibnamefont{Street}},
  \bibinfo{journal}{J. Appl. Phys.} \textbf{\bibinfo{volume}{85}},
  \bibinfo{pages}{7833} (\bibinfo{year}{1999}).

\bibitem[{\citenamefont{Govor}(1986)}]{govor86}
\bibinfo{author}{\bibfnamefont{G.~A.} \bibnamefont{Govor}},
  \bibinfo{journal}{J. of Magnetism and Magnetic Materials}
  \textbf{\bibinfo{volume}{54}}, \bibinfo{pages}{1361} (\bibinfo{year}{1986}).

\bibitem[{\citenamefont{Willis and Rooksby}(1954)}]{willi54}
\bibinfo{author}{\bibfnamefont{B.~T.} \bibnamefont{Willis}} \bibnamefont{and}
  \bibinfo{author}{\bibfnamefont{H.~P.} \bibnamefont{Rooksby}},
  \bibinfo{journal}{Proc. Phys. Soc. London Sect. B}
  \textbf{\bibinfo{volume}{67}}, \bibinfo{pages}{290} (\bibinfo{year}{1954}).

\bibitem[{\citenamefont{Wilson and Kasper}(1964)}]{wilson64}
\bibinfo{author}{\bibfnamefont{R.~H.} \bibnamefont{Wilson}} \bibnamefont{and}
  \bibinfo{author}{\bibfnamefont{J.~S.} \bibnamefont{Kasper}},
  \bibinfo{journal}{Acta Cryst.} \textbf{\bibinfo{volume}{17}},
  \bibinfo{pages}{95} (\bibinfo{year}{1964}).

\bibitem[{\citenamefont{Tanaka et~al.}(1994)\citenamefont{Tanaka, Harbison,
  Park, Park, Shin, and Rothberg}}]{tanaka94}
\bibinfo{author}{\bibfnamefont{M.}~\bibnamefont{Tanaka}},
  \bibinfo{author}{\bibfnamefont{J.}~\bibnamefont{Harbison}},
  \bibinfo{author}{\bibfnamefont{M.~C.} \bibnamefont{Park}},
  \bibinfo{author}{\bibfnamefont{Y.~S.} \bibnamefont{Park}},
  \bibinfo{author}{\bibfnamefont{T.}~\bibnamefont{Shin}}, \bibnamefont{and}
  \bibinfo{author}{\bibfnamefont{G.~M.} \bibnamefont{Rothberg}},
  \bibinfo{journal}{J. Appl. Phys.} \textbf{\bibinfo{volume}{76}},
  \bibinfo{pages}{6278} (\bibinfo{year}{1994}).

\bibitem[{\citenamefont{Schippan et~al.}(1999)\citenamefont{Schippan, Trampert,
  D\"aweritz, and Ploog}}]{schippan99}
\bibinfo{author}{\bibfnamefont{F.}~\bibnamefont{Schippan}},
  \bibinfo{author}{\bibfnamefont{A.}~\bibnamefont{Trampert}},
  \bibinfo{author}{\bibfnamefont{L.}~\bibnamefont{D\"aweritz}},
  \bibnamefont{and} \bibinfo{author}{\bibfnamefont{K.~H.} \bibnamefont{Ploog}},
  \bibinfo{journal}{J. Vac. Sci. Technol. B} \textbf{\bibinfo{volume}{17}},
  \bibinfo{pages}{1716} (\bibinfo{year}{1999}).

\bibitem[{\citenamefont{Kaganer et~al.}(2000)\citenamefont{Kaganer, Jenichen,
  Schippan, Braun, D\"aweritz, and Ploog}}]{kag00}
\bibinfo{author}{\bibfnamefont{V.~M.} \bibnamefont{Kaganer}},
  \bibinfo{author}{\bibfnamefont{B.}~\bibnamefont{Jenichen}},
  \bibinfo{author}{\bibfnamefont{F.}~\bibnamefont{Schippan}},
  \bibinfo{author}{\bibfnamefont{W.}~\bibnamefont{Braun}},
  \bibinfo{author}{\bibfnamefont{L.}~\bibnamefont{D\"aweritz}},
  \bibnamefont{and} \bibinfo{author}{\bibfnamefont{K.~H.} \bibnamefont{Ploog}},
  \bibinfo{journal}{Phys. Rev. Lett.} \textbf{\bibinfo{volume}{85}},
  \bibinfo{pages}{341} (\bibinfo{year}{2000}).

\bibitem[{\citenamefont{Kaganer et~al.}(2002)\citenamefont{Kaganer, Jenichen,
  Schippan, Braun, D\"aweritz, and Ploog}}]{kag02}
\bibinfo{author}{\bibfnamefont{V.~M.} \bibnamefont{Kaganer}},
  \bibinfo{author}{\bibfnamefont{B.}~\bibnamefont{Jenichen}},
  \bibinfo{author}{\bibfnamefont{F.}~\bibnamefont{Schippan}},
  \bibinfo{author}{\bibfnamefont{W.}~\bibnamefont{Braun}},
  \bibinfo{author}{\bibfnamefont{L.}~\bibnamefont{D\"aweritz}},
  \bibnamefont{and} \bibinfo{author}{\bibfnamefont{K.~H.} \bibnamefont{Ploog}},
  \bibinfo{journal}{Phys. Rev. B} \textbf{\bibinfo{volume}{66}},
  \bibinfo{pages}{045305} (\bibinfo{year}{2002}).

\bibitem[{\citenamefont{Plake et~al.}(2002)\citenamefont{Plake, Ramsteiner,
  Kaganer, Jenichen, K\"astner, , D\"aweritz, and Ploog}}]{plake02}
\bibinfo{author}{\bibfnamefont{T.}~\bibnamefont{Plake}},
  \bibinfo{author}{\bibfnamefont{M.}~\bibnamefont{Ramsteiner}},
  \bibinfo{author}{\bibfnamefont{V.~M.} \bibnamefont{Kaganer}},
  \bibinfo{author}{\bibfnamefont{B.}~\bibnamefont{Jenichen}},
  \bibinfo{author}{\bibfnamefont{M.}~\bibnamefont{K\"astner}}, ,
  \bibinfo{author}{\bibfnamefont{L.}~\bibnamefont{D\"aweritz}},
  \bibnamefont{and} \bibinfo{author}{\bibfnamefont{K.~H.} \bibnamefont{Ploog}},
  \bibinfo{journal}{Appl. Phys. Lett.} \textbf{\bibinfo{volume}{80}},
  \bibinfo{pages}{2523} (\bibinfo{year}{2002}).

\bibitem[{\citenamefont{Jenichen et~al.}(2004)\citenamefont{Jenichen, Kaganer,
  Herrmann, Wan, D\"aweritz, and Ploog}}]{jeni04}
\bibinfo{author}{\bibfnamefont{B.}~\bibnamefont{Jenichen}},
  \bibinfo{author}{\bibfnamefont{V.~M.} \bibnamefont{Kaganer}},
  \bibinfo{author}{\bibfnamefont{C.}~\bibnamefont{Herrmann}},
  \bibinfo{author}{\bibfnamefont{L.}~\bibnamefont{Wan}},
  \bibinfo{author}{\bibfnamefont{L.}~\bibnamefont{D\"aweritz}},
  \bibnamefont{and} \bibinfo{author}{\bibfnamefont{K.~H.} \bibnamefont{Ploog}},
  \bibinfo{journal}{Z. Kristallogr.} \textbf{\bibinfo{volume}{219}},
  \bibinfo{pages}{201} (\bibinfo{year}{2004}).

\bibitem[{\citenamefont{Bean and Rodbell}(1962)}]{bean62}
\bibinfo{author}{\bibfnamefont{C.~P.} \bibnamefont{Bean}} \bibnamefont{and}
  \bibinfo{author}{\bibfnamefont{D.~S.} \bibnamefont{Rodbell}},
  \bibinfo{journal}{Phys. Rev.} \textbf{\bibinfo{volume}{126}},
  \bibinfo{pages}{104} (\bibinfo{year}{1962}).

\bibitem[{\citenamefont{Menyuk et~al.}(1969)\citenamefont{Menyuk, Kafalas,
  Dwight, and Goodenough}}]{menyuk69}
\bibinfo{author}{\bibfnamefont{N.}~\bibnamefont{Menyuk}},
  \bibinfo{author}{\bibfnamefont{J.~A.} \bibnamefont{Kafalas}},
  \bibinfo{author}{\bibfnamefont{K.}~\bibnamefont{Dwight}}, \bibnamefont{and}
  \bibinfo{author}{\bibfnamefont{J.~B.} \bibnamefont{Goodenough}},
  \bibinfo{journal}{Phys. Rev.} \textbf{\bibinfo{volume}{177}},
  \bibinfo{pages}{942} (\bibinfo{year}{1969}).

\bibitem[{\citenamefont{Iikawa et~al.}(2005)\citenamefont{Iikawa, Brasil,
  Adriano, Couto, Giles, Santos, D\"aweritz, Rungger, and Sanvito}}]{iikawa05}
\bibinfo{author}{\bibfnamefont{F.}~\bibnamefont{Iikawa}},
  \bibinfo{author}{\bibfnamefont{M.~J.~S.} \bibnamefont{Brasil}},
  \bibinfo{author}{\bibfnamefont{C.}~\bibnamefont{Adriano}},
  \bibinfo{author}{\bibfnamefont{O.~D.~D.} \bibnamefont{Couto}},
  \bibinfo{author}{\bibfnamefont{C.}~\bibnamefont{Giles}},
  \bibinfo{author}{\bibfnamefont{P.~V.} \bibnamefont{Santos}},
  \bibinfo{author}{\bibfnamefont{L.}~\bibnamefont{D\"aweritz}},
  \bibinfo{author}{\bibfnamefont{I.}~\bibnamefont{Rungger}}, \bibnamefont{and}
  \bibinfo{author}{\bibfnamefont{S.}~\bibnamefont{Sanvito}},
  \bibinfo{journal}{Phys. Rev. Lett.} \textbf{\bibinfo{volume}{95}},
  \bibinfo{pages}{077203} (\bibinfo{year}{2005}).

\bibitem[{\citenamefont{Takagagi et~al.}(2006)\citenamefont{Takagagi, Jenichen,
  Herrmann, Wiebecke, D\"aweritz, and Ploog}}]{taka05}
\bibinfo{author}{\bibfnamefont{Y.}~\bibnamefont{Takagagi}},
  \bibinfo{author}{\bibfnamefont{B.}~\bibnamefont{Jenichen}},
  \bibinfo{author}{\bibfnamefont{C.}~\bibnamefont{Herrmann}},
  \bibinfo{author}{\bibfnamefont{E.}~\bibnamefont{Wiebecke}},
  \bibinfo{author}{\bibfnamefont{L.}~\bibnamefont{D\"aweritz}},
  \bibnamefont{and} \bibinfo{author}{\bibfnamefont{K.~H.} \bibnamefont{Ploog}},
  \bibinfo{journal}{Phys. Rev. B} \textbf{\bibinfo{volume}{73}},
  \bibinfo{pages}{125324} (\bibinfo{year}{2006}).

\bibitem[{\citenamefont{K\"astner et~al.}(2000)\citenamefont{K\"astner,
  Schippan, Sch\"utzend\"ube, D\"aweritz, and Ploog}}]{kaest00}
\bibinfo{author}{\bibfnamefont{M.}~\bibnamefont{K\"astner}},
  \bibinfo{author}{\bibfnamefont{F.}~\bibnamefont{Schippan}},
  \bibinfo{author}{\bibfnamefont{P.}~\bibnamefont{Sch\"utzend\"ube}},
  \bibinfo{author}{\bibfnamefont{L.}~\bibnamefont{D\"aweritz}},
  \bibnamefont{and} \bibinfo{author}{\bibfnamefont{K.~H.} \bibnamefont{Ploog}},
  \bibinfo{journal}{J. Vac. Sci. Technol. B} \textbf{\bibinfo{volume}{18}},
  \bibinfo{pages}{2052} (\bibinfo{year}{2000}).

\bibitem[{\citenamefont{D\"aweritz et~al.}(2004)\citenamefont{D\"aweritz, Wan,
  Jenichen, Herrmann, Mohanty, Trampert, and Ploog}}]{daw04}
\bibinfo{author}{\bibfnamefont{L.}~\bibnamefont{D\"aweritz}},
  \bibinfo{author}{\bibfnamefont{L.}~\bibnamefont{Wan}},
  \bibinfo{author}{\bibfnamefont{B.}~\bibnamefont{Jenichen}},
  \bibinfo{author}{\bibfnamefont{C.}~\bibnamefont{Herrmann}},
  \bibinfo{author}{\bibfnamefont{J.}~\bibnamefont{Mohanty}},
  \bibinfo{author}{\bibfnamefont{A.}~\bibnamefont{Trampert}}, \bibnamefont{and}
  \bibinfo{author}{\bibfnamefont{K.~H.} \bibnamefont{Ploog}},
  \bibinfo{journal}{J. Appl. Phys.} \textbf{\bibinfo{volume}{96}},
  \bibinfo{pages}{5056} (\bibinfo{year}{2004}).

\bibitem[{\citenamefont{Jenichen et~al.}(2002)\citenamefont{Jenichen, Kaganer,
  Schippan, Braun, D\"aweritz, and Ploog}}]{jeni02}
\bibinfo{author}{\bibfnamefont{B.}~\bibnamefont{Jenichen}},
  \bibinfo{author}{\bibfnamefont{V.~M.} \bibnamefont{Kaganer}},
  \bibinfo{author}{\bibfnamefont{F.}~\bibnamefont{Schippan}},
  \bibinfo{author}{\bibfnamefont{W.}~\bibnamefont{Braun}},
  \bibinfo{author}{\bibfnamefont{L.}~\bibnamefont{D\"aweritz}},
  \bibnamefont{and} \bibinfo{author}{\bibfnamefont{K.~H.} \bibnamefont{Ploog}},
  \bibinfo{journal}{Mat. Science and Eng. B} \textbf{\bibinfo{volume}{91}},
  \bibinfo{pages}{433} (\bibinfo{year}{2002}).

\bibitem[{\citenamefont{Jenichen et~al.}(1993)\citenamefont{Jenichen, Brandt,
  and Ploog}}]{jeni93}
\bibinfo{author}{\bibfnamefont{B.}~\bibnamefont{Jenichen}},
  \bibinfo{author}{\bibfnamefont{O.}~\bibnamefont{Brandt}}, \bibnamefont{and}
  \bibinfo{author}{\bibfnamefont{K.~H.} \bibnamefont{Ploog}},
  \bibinfo{journal}{Appl. Phys. Lett.} \textbf{\bibinfo{volume}{63}},
  \bibinfo{pages}{156} (\bibinfo{year}{1993}).

\bibitem[{\citenamefont{Landau and Lifschitz}(1975)}]{landau75}
\bibinfo{author}{\bibfnamefont{L.~D.} \bibnamefont{Landau}} \bibnamefont{and}
  \bibinfo{author}{\bibfnamefont{I.~M.} \bibnamefont{Lifschitz}},
  \emph{\bibinfo{title}{Statistische Physik}}
  (\bibinfo{publisher}{Akademie-Verlag}, \bibinfo{address}{Berlin, Germany},
  \bibinfo{year}{1975}).

\bibitem[{\citenamefont{Wilke and Bohm}(1988)}]{bohm88}
\bibinfo{author}{\bibfnamefont{K.~T.} \bibnamefont{Wilke}} \bibnamefont{and}
  \bibinfo{author}{\bibfnamefont{J.}~\bibnamefont{Bohm}},
  \emph{\bibinfo{title}{Kristallz\"uchtung}} (\bibinfo{publisher}{Deutscher
  Verlag der Wissenschaften}, \bibinfo{address}{Berlin, Germany},
  \bibinfo{year}{1988}).

\end{thebibliography}
\newpage

\section*{Figures}
\vspace{5cm}
\newpage

%%\newpage

%%\newpage

%%\newpage

%%\newpage

\end{document}